\def\imat  {{\rm i}}
\documentstyle[twocolumn,floats,epsf,ifthen,aps,prb]{revtex} 

\begin{document} 

 

\title{Nonideal quantum detectors in Bayesian formalism} 

\author{Alexander N. Korotkov} 
\address{ 
Department of Electrical Engineering, University of California, 
Riverside, CA 92521-0204. 
} 
\date{\today} 
 
\maketitle 
 
\begin{abstract} 
The Bayesian formalism for a continuous measurement of  solid-state qubits 
is derived for a model which takes into account several factors of the 
detector nonideality. In particular, we consider additional classical output 
and backaction noises (with finite correlation), together with 
quantum-limited output and backaction noises, and take into account possible 
asymmetry of the detector coupling. The formalism is first derived for a 
single qubit and then generalized to the measurement of entangled qubits. 
\end{abstract} 
 
\narrowtext 
 
\vspace{0.6cm} 

\section{Introduction} 

       The problem of continuous qubit measurement is of a significant 
importance for solid-state quantum computing\cite{Nielsen} because
the measurement of a solid-state qubit typically requires a significant
time and thus can interplay nontrivially
\cite{Gurvitz,Makhlin,Kor-99,Averin-res,Goan} 
with the intrinsic evolution of the qubit system. 
The evolution of a single solid-state qubit (without ensemble averaging) 
due to continuous measurement can be described by 
the Bayesian formalism (for review see Ref.\ \cite{Kor-rev2}) which
takes into account the noisy measurement output of the detector. 
The Bayesian formalism practically coincides with the version of the 
quantum trajectory formalism \cite{Goan} adapted to solid-state setups 
from the theory developed for quantum optics. \cite{Wiseman-93} 

   One of the main predictions of the Bayesian formalism is the absence 
of the single qubit decoherence during the measurement by a good
(ideal) detector, \cite{Kor-99}
in contrast to decoherence of an ensemble of 
qubits.\cite{Caldeira,Zurek} Moreover, the 
state of a solid-state qubit can be gradually purified due to 
continuous measurement. In particular, 
this makes possible to monitor the phase of quantum coherent (Rabi) 
oscillations of the qubit. Such monitoring can be naturally used
in the quantum feedback control \cite{Kor-rev1,Ruskov-fb} of the Rabi 
oscillations which suppresses the qubit decoherence due to environment
(for quantum feedback in quantum optics see, e.g., Refs.\ 
\cite{Wiseman-93-fb,Tombesi,Doherty,Armen}). 
Another potentially useful application of the Bayesian formalism is
a recent prediction that two qubits can be made fully entangled by
their continuous measurement by an equally coupled detector.\cite{Ruskov-ent}

    The efficiency of the quantum feedback loop operation crucially  
depends on the  ideality (quantum efficiency) of the detector. For example,
100\% synchronization between the qubit Rabi oscillations and desired
pure oscillations is possible only for 100\% ideal detector.\cite{Ruskov-fb}
Many other effects related to continuous measurement of solid-state qubits,
which have been predicted using the Bayesian formalism (see, e.g., Refs.\
\cite{Kor-99,Ruskov-ent,Kor-sp,Kor-exp,Goan2}) also depend significantly 
on the detector ideality. The ideality $\eta$ of a continuously operating
solid-state detector can be generally defined as 
a ratio between the detector performance and the performance of a 
quantum-limited detector, in which the output and backaction noises 
are strictly related by the lower bound of an inequality similar
to the Heisenberg uncertainty relation. More exact definition will be
discussed later. 

     A Quantum Point Contact (QPC) at low temperature is theoretically an 
ideal quantum detector \cite{Kor-99} that follows from the results of 
Refs.\ \cite{Gurvitz,Aleiner}. A nearly ideal operation of the QPC 
has been demonstrated experimentally.\cite{Buks,Sprinzak} 
The fact that a SQUID can theoretically reach the limit of an ideal detector
follows \cite{Averin-book} 
from the results of Ref.\ \cite{Danilov}. A normal state single-electron 
transistor (SET) is not a good quantum detector 
at usual operating points above the Coulomb Blockade 
threshold \cite{Makhlin,Kor-rev1}. However, its quantum efficiency 
improves when we go closer to the threshold \cite{Kor-rev1,Devoret} 
and becomes much better when the operating point is in the cotunneling
range (below the threshold), in which case the limit of an ideal
detector can be achieved \cite{Averin-cotun,Brink}. Superconducting 
SET is generally better than normal SET as a quantum-limited detector 
and can approach 100\% ideality in the supercurrent regime \cite{Zorin} 
as well as in the double Josephson-plus-quasiparticle regime \cite{Clerk}.
Finally, the resonant-tunneling SET \cite{Averin-res} can reach  
complete ideality in the small-bias limit. 

        In the simplest version of the Bayesian formalism \cite{Kor-99} 
a nonideal solid-state detector is modeled as an ideal symmetrically 
coupled detector \cite{symmetric}
and a ``pure dephasor'' in parallel (environment or just extra backaction 
noise). In this case the nonideality leads to an extra term in the Bayesian 
equations, which introduces the gradual decay of the nondiagonal elements 
of the density matrix of the measured qubit. It was implied that such 
backaction dephasing 
is also equivalent to the extra noise at the detector output. However, the
equivalence has never been proven explicitly, and this is one of the goals 
of the present paper. 

        In a more advanced version of the Bayesian formalism,  
\cite{Kor-LT,Kor-rev1} a possible correlation between the output noise 
of a nonideal detector and the backaction noise is taken into account. 
However, the formulas 
for the evolution of the qubit density matrix in this case have been
presented without any derivation, just from physical intuition. 
Moreover, comparison of these formulas with the results of Ref.\ \cite{Goan2}
for an ideal but asymmetrically coupled detector (which  shifts
the energy levels 
of the measured qubit) reveals some difference. Even though the difference
is minor (second order in the detector response, which is assumed to be 
small), it points to some incorrectness of the initial formulas of Ref.\ 
\cite{Kor-LT} (corrected formulas can be found in Ref.\ \cite{Kor-rev2}). 
The main goal of this paper is to present a mathematical derivation 
of the Bayesian formalism for a nonideal detector with correlated output
and backaction noises, using the phenomenological model which adds correlated
classical noise to the quantum noise of an asymmetric ideal detector.
We start with the measurement of one qubit and then generalize the 
formalism to the continuous measurement of an arbitrary number of entangled
qubits. 

     Notice that the issue of the asymmetric detector coupling 
to qubit has been recently discussed for a QPC in terms of the tunneling 
phase control by the qubit state.
\cite{Sprinzak,Averin-book,Stodolsky,Kor-Av} 
For a small-transparency QPC the formalism is significantly simplified 
\cite{Goan2} and is a direct generalization of the model of Ref.\ 
\cite{Gurvitz}. We will use the results of Ref.\ \cite{Goan2} to model
an ideal asymmetric detector. 

        While we model the detector nonideality by an additional classical
noise, let us mention a different approach to the nonideality in Ref.\
\cite{Goan2}, in which a random fraction of electrons tunneled through 
the detector is assumed to be missing. In our opinion, such model is not 
well applicable to solid-state detectors, even though it perfectly
describes the inefficiency of a photodetector in a similar problem 
in quantum optics.

  \section{Model}

\begin{figure} 
\centerline{
\epsfxsize=3.1in 
\vspace{0.1cm}
\epsfbox{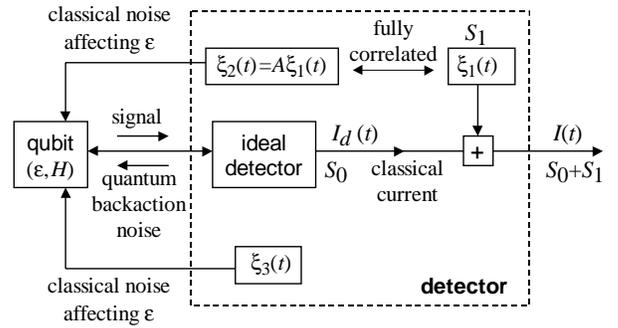}
} 
\vspace{0.3cm} 
\caption{Schematic of a nonideal solid state detector measuring a qubit
state. The detector is modeled as an ideal quantum detector and three
sources of additional classical noise: output noise $\xi_1(t)$ with
the white spectral density $S_1$, backaction noise $\xi_2(t)$ fully
correlated with $\xi_1(t)$, and uncorrelated backaction noise
$\xi_3(t)$. The total noise density $S_0+S_1$ of the output detector 
current $I(t)$ includes the contribution $S_0$ from the noise of 
ideal detector current $I_d(t)$. 
 }
\label{Fig1}\end{figure}

    We will use the phenomenological model of a nonideal solid-state detector 
of a qubit state shown in Fig.\ \ref{Fig1}. It consists of an ideal detector 
and three sources of additional classical noise. We assume that the detector
output is the noisy current $I(t)$ (we have in mind a QPC or a 
SET as a detector). 
 The ideal detector is characterized by the output noise spectral density 
$S_0$ [we assume flat (``white'') noise spectrum] and its backaction onto
the measured qubit which will be called ``quantum noise''. (Actually, 
because of the quantum relation between the output noise and the backaction,
the output noise could also be called quantum; however, we will avoid such
terminology, emphasizing the assumption 
that the quantum behavior does not propagate beyond the 
ideal detector.)

        The first source of an additional classical noise adds the 
noisy component $\xi_1(t)$ with the white spectral density $S_1$ to the 
output $I_d(t)$ of the ideal detector, so that the final output is
$I(t)=I_d(t)+\xi_1(t)$. The second noise source is the classical noise
$\xi_2(t)$ which is 100\% correlated with (proportional to) the noise 
$\xi_1(t)$ and affects the qubit energy asymmetry $\varepsilon$. 
[The qubit Hamiltonian is
        \begin{equation}
{\cal H}_{qb} = \frac{\varepsilon}{2}\, (c_2^\dagger  c_2 - c_1^\dagger c_1) 
        + H \, (c_1^\dagger c_2 +c_2^\dagger c_1) ,
        \end{equation}
where the tunneling strength $H$ is assumed to be real without loss of 
generality.] 
The relative magnitude of the noise $\xi_2(t)=A\xi_1(t)$ is characterized
by the parameter $A$. Finally, the third 
classical noise source is the white noise $\xi_3(t)$ which also affects
the qubit energy asymmetry $\varepsilon$
[so that $\varepsilon \rightarrow \varepsilon +\xi_2(t) +\xi_3 (t)$]. 
The second and third noise sources together are obviously equivalent 
to one white noise source, partially correlated
with $\xi_1(t)$. However, we prefer to split it into the fully correlated
and uncorrelated parts for clarity. Obviously, the qubit parameter $H$ 
can also be affected by the detector noise; however, we do not take this
effect into account, because the qubit dephasing is more naturally caused
by the noise of its energy asymmetry $\varepsilon$ (which corresponds to the
measured degree of freedom) and also because the induced noise of $H$ 
is negligible, for example, for a single-Cooper-pair qubit measured by 
an SET. 

        Let us start with the symmetric ideal detector,
neglect all classical noises $\xi_{1,2,3}(t)$ and use the basic Bayesian
formalism to describe the measurement process (i.e.\ 
the result of quantum backaction onto qubit); 
then the evolution of the qubit density matrix $\rho_{ij}(t)$ is 
\cite{Kor-99,Kor-rev2,Kor-rev1} 
        \begin{eqnarray}
&&  \dot{\rho}_{11}=  -\dot{\rho}_{22}=  
-2\,\frac{H}{\hbar}\,\mbox{Im}\,\rho_{12}
         +\rho_{11}\rho_{22}\, \frac{2\Delta I}{S_0}\, [I_d(t)-I_0], 
        \label{Bayes1}\\ 
&&  {\dot\rho}_{12}=  \imat\, \frac{\varepsilon}{\hbar }\,
        \rho_{12}+ 
        \imat \, \frac{H}{\hbar } \, (\rho_{11}-\rho_{22})
\nonumber \\
&& \hspace{1cm}  -( \rho_{11}-  \rho_{22})  \frac{\Delta I}{S_0} \, 
[I_d(t)-I_0]\, \rho_{12} .
        \label{Bayes2}  \end{eqnarray}
Here $\Delta I \equiv I_1 -I_2$ is the detector response, $I_1$ is the average
detector current for the qubit state $|1\rangle$, $I_2$ is the 
average current for the state $|2\rangle$, and $I_0\equiv (I_1+I_2)/2$.
For the validity of Eqs.\ (\ref{Bayes1})--(\ref{Bayes2}) we have to assume
\cite{Kor-99,Kor-rev2,Kor-rev1} the weakly responding detector, 
$|\Delta I|\ll I_0$, sufficiently large detector voltage (much larger than
the qubit energies), and assume that the passage of individual electrons 
in the detector is much faster than the qubit evolution, $I_0/e \gg 
(4H^2+\varepsilon^2)/\hbar  $, so that the current can be considered 
as continuous.

        For simulations, Eqs.\ (\ref{Bayes1})--(\ref{Bayes2}) should
be complemented by the equation
        \begin{equation} 
I_d(t) -I_0 = \frac{\Delta I}{2}\, (\rho_{11} -\rho_{22}) +\xi_0 (t) ,
        \label{I(t)}\end{equation}
where $\xi_0(t)$ is the pure output noise of the ideal detector with
flat spectral density $S_0$. 

        Equations (\ref{Bayes1})--(\ref{Bayes2})  are written in the
so-called Stratonovich form,\cite{Oksendal} which assumes symmetric
definition of the derivative, $\dot\rho (t)=\lim_{\tau \rightarrow 0}
[\rho (t+\tau /2) -\rho (t-\tau /2)]/\tau$. 
For the forward definition of derivative, 
$\dot\rho (t)=\lim_{\tau \rightarrow 0}[\rho (t+\tau ) -\rho (t)]/\tau$
(It\^o form), the Eqs.\ (\ref{Bayes1})--(\ref{Bayes2}) transform into 
\cite{Kor-rev1} 
        \begin{eqnarray}
&& 
   \dot{\rho}_{11}  =   -\dot{\rho}_{22}=  -2\,\frac{H}{\hbar}\,\mbox{Im}\,
          \rho_{12}
         +\rho_{11}\rho_{22}\, \frac{2\Delta I}{S_0}\, \xi_0 (t)\, ,  
        \label{Ito1}\\ 
&&
 {\dot\rho}_{12}  =   \imat\, \frac{\varepsilon}{\hbar }\,\rho_{12} +   
        \imat \, \frac{H}{\hbar } \, (\rho_{11}-\rho_{22})
\nonumber\\
&& \hspace{1cm}
 -  ( \rho_{11}-  \rho_{22})  \frac{\Delta I}{S_0} \, 
\rho_{12}\, \xi_0 (t) - \frac{(\Delta I)^2}{4S_0}  \, 
\rho_{12} \, , 
        \label{Ito2} 
        \end{eqnarray}
while the relation (\ref{I(t)}) remains unchanged.
[The general rule of transformation is the following:\cite{Oksendal} 
for an arbitrary 
system of equations $\dot x_i(t)=G_i({\bf x},t)+F_i({\bf x},t)\xi(t)$
in the Stratonovich form, the corresponding equations in the It\^o form
are $\dot x_i(t)=G_i({\bf x},t)+F_i({\bf x},t)\xi(t)+
(S_\xi /4)\sum_k F_k({\bf x},t)\partial F_i({\bf x},t)/\partial x_k$.]
 The advantage of 
the It\^o form is that the averaging over the noise $\xi_0(t)$ is
straightforward (we just need to eliminate terms with $\xi_0$), so it is
easy to obtain the ensemble averaged evolution:
        \begin{eqnarray}
&& 
   \dot{\rho}_{11}  =   -\dot{\rho}_{22}=  -2\,\frac{H}{\hbar}\,\mbox{Im}\,
          \rho_{12} \, ,  
        \label{aver1}\\ 
&&
 {\dot\rho}_{12}  =   \imat\, \frac{\varepsilon}{\hbar }\,\rho_{12} +   
        \imat \, \frac{H}{\hbar } \, (\rho_{11}-\rho_{22})
 - \frac{(\Delta I)^2}{4S_0}  \, 
\rho_{12} \, .
        \label{aver2} 
        \end{eqnarray}
On the other hand, the advantage of the Stratonovich form is the validity
of usual calculus rules (which do not work in the It\^o form) and therefore
easier physical interpretation of equations.

        Let us emphasize that the single qubit in this model (which
assumes ideal detector) 
does not decohere during the measurement process, as easier to see from 
Eqs.\  (\ref{Bayes1})--(\ref{Bayes2}). However, because of the probabilistic
nature of quantum measurement, the ensemble of qubits does decohere
(different qubits will go along different ``trajectories''). The 
ensemble decoherence rate $(\Delta I)^2/4S_0$ is determined by this quantum
randomness and therefore can be naturally called ``quantum-limited''
decoherence rate. (It can also be called ``information-limited'' ensemble 
decoherence, since its origin is the tendency of qubit state to evolve 
either into state $|1\rangle$ or $|2\rangle$, corresponding to the 
information acquired from the measurement.\cite{Kor-rev1} )

        While it is not trivial to take into account additional 
classical noises $\xi_1$ and $\xi_2$ (this will be done in
the following Sections), the account of the noise $\xi_3$ is very simple.
It leads to the additional dephasing term $-\gamma_3 \rho_{12}$ in
Eqs.\ (\ref{Bayes2}), (\ref{Ito2}), and (\ref{aver2}), where 
$\gamma_3=S_3/4\hbar^2$ 
is proportional to the spectral density $S_3$ of $\xi_3(t)$.
We will characterize this noise by the dephasing rate $\gamma_3$ instead of
characterizing it by $S_3$. (The relation $\gamma_3=S_3/4\hbar^2$ 
can be easily derived adding $\xi_3(t)$ to $\varepsilon$ in the Stratonovich
form, then translating the equation into It\^o form, and averaging over 
$\xi_3$.)

        The natural definition of the detector ideality factor 
$\eta$ in this case is \cite{Kor-99,Kor-rev2,Kor-rev1} 
\begin{equation}
\eta \equiv 
\frac{\Gamma_0}{\Gamma_\Sigma} , 
\end{equation}
where $\Gamma_0 \equiv (\Delta I)^2/4S_0$ is the quantum-limited contribution
and  $\Gamma_\Sigma =\Gamma_0+ \gamma_3$ is the total ensemble 
dephasing rate. Simply speaking, this definition of ideality is 
the ratio between quantum contribution and total backaction noise.

\section{Ideal symmetric detector and additional output noise} 

        Let us now take into account additional output noise $\xi_1(t)$, 
while $\xi_2(t)$ is still zero. We also switch off $\xi_3(t)$, since 
it is trivial to add its effect later. In order to derive Bayesian equations
in this case, let us also assume $H=\varepsilon =0$ (``frozen'' qubit)
and add the effects of $H$ and $\varepsilon$ later. For $H=\varepsilon =0$, 
Eqs.\ (\ref{Bayes1})--(\ref{Bayes2}) have a simple solution
\cite{Kor-99,Kor-rev1} which can be interpreted as a consequence of 
the ``Quantum Bayes theorem'':
\cite{Gardiner}
        \begin{eqnarray}
&&\rho_{11}(\tau )=
\left[ 1+\frac{\rho_{22}(\tau )}{\rho_{11}(\tau )}\right]^{-1} 
\nonumber \\
&&\hspace{1.1cm} 
= \left[ 1+\frac{\rho_{22}(0)}{\rho_{11}(0)}
\frac{\exp [-(\overline{I_d}-I_2)^2\tau /S_0]}
{\exp [-(\overline{I_d}-I_1)^2\tau /S_0]}  \right]^{-1} , 
        \label{QBayes1} \\
&& \rho_{22}(\tau )=1-\rho_{11}(\tau )
        \label{QBayes2} \\
&& \rho_{12}(\tau ) = \rho_{12}(0) 
\frac{[\rho_{11}(\tau )\rho_{22}(\tau )]^{1/2}}
{[\rho_{11}(0)\rho_{22}(0)]^{1/2}}
        \label{QBayes3}\\
&& = \frac{
\rho_{12}(0) \exp [-\frac{(\Delta I)^2\tau}{4S_0}] 
\exp [-\frac{(\overline{I_d}-I_0)^2\tau }{S_0}] 
}{
\rho_{11}(0)\exp [-\frac{(\overline{I_d}-I_1)^2\tau }{S_0}]
+\rho_{22}(0)\exp [-\frac{(\overline{I_d}-I_2)^2\tau }{S_0}]
} .
        \label{QBayes4}\end{eqnarray}
where $\overline{I_d}$ is the average of the detector current during the
time interval between 0 and $\tau$:
\begin{equation} 
     \overline{I_d}=\frac{1}{\tau} \int_{0}^{\tau} I_d (t) dt. 
\end{equation}
Here Eq.\ (\ref{QBayes1}) is the consequence of the classical Bayes theorem
\cite{Bayes} and Eq.\ (\ref{QBayes3}) says that the degree of the qubit 
purity is conserved. [It is easy to include the effect of finite 
$\varepsilon$, 
which just leads to an extra factor $\exp(\imat \varepsilon \tau /\hbar )$ 
in Eqs.\ (\ref{QBayes3}) and (\ref{QBayes4});
however, we will not do that in order to keep the formulas shorter.] 

     Since the detector output is now $I(t) =I_d(t)+\xi_1(t)$, we have
to express $\rho_{ij}(\tau )$ in terms of $I(t)$ and average it over the 
noise $\xi_1(t)$. Naively thinking, we have to use the substitution 
\begin{equation}
\overline{I_d}=\overline{I}-x , \,\,\,  
\overline{I}=\frac{1}{\tau} \int_{0}^{\tau} I (t) dt,  \,\,\, 
x=\frac{1}{\tau}\int_0^\tau \xi_1(t)dt, 
\end{equation}
and average Eqs.\ (\ref{QBayes1}) and (\ref{QBayes4}) over the noise 
contribution $x$ using the weight factor 
$p(x)= (2\pi D_1)^{-1/2}\exp (-x^2/2D_1)$ where 
$D_1=S_1/2\tau$ is the variance of $x$.
However, this is not a correct procedure because
the probability distribution of $x$ is correlated with 
$\overline{I}$ (though it is not correlated with $\overline{I_d}$).
So instead, we have to use the conditional distribution of $x$ for a given
$\overline{I}$: 
\begin{eqnarray}
&& p(x)= P(x)/\int P(x')dx',
\label{p1(x)}\\
&& P(x)=\frac{\exp (-x^2/2D_1) }{\sqrt{2\pi D_1}} \left[
\rho_{11}(0)\frac{\exp [-(\overline{I}-x-I_1)^2/2D_0]}{\sqrt{2\pi D_0}}
\right.
\nonumber\\
&& \hspace{1.0cm} \left.
+\rho_{22}(0)\frac{\exp [-(\overline{I}-x-I_2)^2/2D_0]}{\sqrt{2\pi D_0}}
\right] ,
\label{p2(x)}\end{eqnarray}
where $D_0=S_0/2\tau$. (Let us stress again that both $\overline{I_d}$ 
and $x$ are assumed to be classical quantities, so the derivation 
is not applicable to the case when the detector output $I_d$ is involved
in further quantum interactions.) 
Introducing the weight factor $p(x)$ into Eq.\ (\ref{QBayes1}),
substituting $\overline{I_d}=\overline{I}-x$, and integrating
over $x$, we get the averaged equation 
        \begin{equation}
\rho_{11}(\tau )=
 \left[ 1+\frac{\rho_{22}(0)}{\rho_{11}(0)}
\frac{\exp [-(\overline{I}-I_2)^2\tau /S_\Sigma]}
{\exp [-(\overline{I}-I_1)^2\tau /S_\Sigma]}  \right]^{-1} , 
\label{r11-1}\end{equation}
where $S_\Sigma \equiv  S_0+S_1$ is the total output noise. 
The only difference compared with Eq.\ (\ref{QBayes1}) 
is the change of $\overline{I_d}$ into $\overline{I}$ and change
of $S_0$ into $S_\Sigma$ (this is quite expected since 
$\rho_{11}$ behaves as a classical probability and the classical Bayes 
formula still works). 

        To calculate $\rho_{12}(\tau )$ averaged over the noise $\xi_1(t)$,
we have to do a similar procedure. We multiple Eq.\ (\ref{QBayes4}) by 
the weight factor $p(x)$, use substitution $\overline{I_d}=\overline{I}-x$, 
and integrate over $x$. In this way we obtain 
\begin{equation}
\rho_{12}(\tau )=\frac{
\rho_{12}(0) \exp [-\frac{(\Delta I)^2 \tau }{4S_0}] 
\exp [-\frac{(\overline{I}-I_0)^2\tau }{S_\Sigma}] 
}{
\rho_{11}(0)\exp [-\frac{(\overline{I}-I_1)^2\tau }{S_\Sigma}]
+\rho_{22}(0)\exp [-\frac{(\overline{I}-I_2)^2\tau }{S_\Sigma}]
} .
\label{r12av1}\end{equation}
Comparing Eqs.\  (\ref{QBayes4}) and (\ref{r12av1}), we see that 
$\overline{I_d}$ changes into $\overline{I}$ and 
$S_0$ changes into $S_\Sigma$, except in the second factor of the numerator,
where $S_0$ remains unchanged. Let us represent this factor as 
$\exp[-(\Delta I)^2\tau /4S_\Sigma]\exp(-\gamma_1 \tau )$, where 
\begin{equation}
\gamma_1=\frac{(\Delta I)^2S_1}{4S_0S_\Sigma} \, .
\label{gamma1}\end{equation} 

So, the effect of additional output noise $\xi_1$ on the Eqs.\  
(\ref{QBayes1})--(\ref{QBayes4}) is the following: the output current
$I_d$ from the ideal part of the detector changes into the output current
$I$, the spectral density $S_0$ corresponding to the ideal part of the
detector changes into the total output noise $S_\Sigma$, and the nondiagonal
matrix element acquires the dephasing factor $\exp (-\gamma_1 \tau)$.
Differentiating the new equations over time (if we do it in a simple 
first-order way, we automatically get equations in the Stratonovich form)
 and adding terms due to $H$, 
$\varepsilon$, and noise $\xi_3$, we obtain  
        \begin{eqnarray}
&&  \dot{\rho}_{11}=  -\dot{\rho}_{22}=  
-2\,\frac{H}{\hbar}\,\mbox{Im}\,\rho_{12}
         +\rho_{11}\rho_{22}\, \frac{2\Delta I}{S_\Sigma}\, [I(t)-I_0], 
        \label{Bayes1-1}\\ 
&&  {\dot\rho}_{12}=  \imat\, \frac{\varepsilon}{\hbar }\,
        \rho_{12}+ 
        \imat \, \frac{H}{\hbar } \, (\rho_{11}-\rho_{22})
\nonumber \\
&& \hspace{0.6cm}  -( \rho_{11}-  \rho_{22})  \frac{\Delta I}{S_\Sigma} \, 
[I(t)-I_0]\, \rho_{12} - (\gamma_1 +\gamma_3) \rho_{12} . 
        \label{Bayes2-1}\end{eqnarray}
We see that the effect of the extra output noise $\xi_{1}(t)$ is similar to 
the effect of the extra backaction noise $\xi_3(t)$ and both lead to the
qubit dephasing.

        From physical reasoning, the way of separation of the detector 
into the ideal part and additional noise sources $\xi_1(t)$ and $\xi_3(t)$
is arbitrary, as long as the total output noise $S_\Sigma$ and total 
ensemble qubit dephasing rate $\Gamma_\Sigma$ are fixed (in other words, 
$S_\Sigma$ and $\Gamma_\Sigma$ are the only physically relevant quantities).
It is easy to check that Eqs.\ (\ref{Bayes1-1})--(\ref{Bayes2-1})
satisfy this requirement because 
\begin{equation}
\gamma_1+\gamma_3  = \Gamma_\Sigma - \frac{(\Delta I)^2}{4S_\Sigma} \, , 
\,\,\,\,\, \Gamma_\Sigma =\frac{(\Delta I)^2}{4S_0}+\gamma_3.
\end{equation}
[The total ensemble dephasing rate $\Gamma_\Sigma$ can be formally 
found from Eqs.\ 
(\ref{Bayes1-1})--(\ref{Bayes2-1}) by translating them into It\^o form
that adds ensemble dephasing rate $(\Delta I)^2/4S_\Sigma)$.]

Comparing Eqs.\ (\ref{Bayes1-1})--(\ref{Bayes2-1}) with Eqs.\
(\ref{Bayes1})--(\ref{Bayes2}), we naturally introduce a more general 
definition of the detector ideality: 
\begin{equation}
\eta =\frac{(\Delta I)^2/4S_\Sigma}{\Gamma_\Sigma}, 
\label{eta-gen}\end{equation}
which is again the ratio of the quantum-limited part of the ensemble qubit
dephasing and its total dephasing rate.
(Notice that the numerator is not the ``real'' quantum backaction  
determined  by  $S_0$,  but  the   information-limited   backaction 
determined by $S_\Sigma$.) 
In particular, for our model in the case $\xi_3(t)=0$ 
(no classical backaction) we obtain $\eta = S_0/(S_0+S_1)$. 

\section{Correlated output and backaction noises}

        Now let us add the classical backaction noise $\xi_2(t)$ which 
affects the qubit energy $\varepsilon$ [so that $\varepsilon \rightarrow 
\varepsilon +\xi_2 (t)$], and which is 
100\% correlated with the output noise source, $\xi_2(t)=A \xi_1(t)$. 
We again start with Eqs.\ (\ref{QBayes1})--(\ref{QBayes4}) for the
``frozen'' qubit. Addition of $\xi_2(t)$ does not affect diagonal
matrix elements, so Eq.\ (\ref{r11-1}) remains unchanged. 
In order to calculate $\rho_{12}(\tau )$ averaged over the noises 
$\xi_1$ and $\xi_2$, we multiply Eq.\ (\ref{QBayes4}) by
the factor $\exp [\imat \hbar^{-1} \int_0^\tau \xi_2 (t) dt] =
\exp (\imat Ax\tau /\hbar )$ and average it over $x$ with the weight $p(x)$ 
given by Eqs.\ (\ref{p1(x)})--(\ref{p2(x)}). This leads to the equation
\begin{eqnarray}
&& \rho_{12}(\tau )=\frac{
\rho_{12}(0) \exp [-\frac{(\Delta I)^2 \tau }{4S_0}] 
\exp [-\frac{(\overline{I}-I_0)^2\tau }{S_\Sigma}] 
}{
\rho_{11}(0)\exp [-\frac{(\overline{I}-I_1)^2\tau }{S_\Sigma}]
+\rho_{22}(0)\exp (-\frac{(\overline{I}-I_2)^2\tau }{S_\Sigma})
} 
\nonumber \\
&&\hspace{1.1cm} \times
\exp[ \frac{\imat}{\hbar} (\overline{I}-I_0) A \frac{S_1}{S_\Sigma} \tau ]
\exp [-\frac{S_0S_1}{S_\Sigma}\frac{A^2}{4\hbar^2}\tau ], 
\label{r12av2}\end{eqnarray}
in which only the second line is different from Eq.\ (\ref{r12av1}).

        Differentiating this equation over time (again,
we automatically obtain the Stratonovich form) and adding terms due to $H$, 
$\varepsilon$ and $\xi_3$, we get 
        \begin{eqnarray}
&& {\dot\rho}_{12}=  \imat\, \frac{\varepsilon}{\hbar }\,
        \rho_{12}+ 
        \imat \, \frac{H}{\hbar } \, (\rho_{11}-\rho_{22})
   +\frac{\imat}{\hbar} [I(t) -I_0] A \frac{S_1}{S_\Sigma} \rho_{12} 
\nonumber \\
&& \hspace{0.2cm}  -( \rho_{11}-  \rho_{22})  \frac{\Delta I}{S_\Sigma} \, 
[I(t)-I_0]\, \rho_{12} - (\gamma_1 +\gamma_2+ \gamma_3) \rho_{12} ,
        \label{Bayes2-2}\end{eqnarray} 
where $\gamma_1$ is given by Eq.\ (\ref{gamma1}) and $\gamma_2$ is 
\begin{equation}
\gamma_2 = \frac{S_0S_1}{S_\Sigma} \frac{A^2}{4\hbar^2}. 
\end{equation}

        Physically relevant parameters of the detector are 
the total output noise $S_\Sigma =S_0+S_1$, total ensemble
dephasing rate $\Gamma_\Sigma$, and the correlation between
output and backaction noises. Since three sources of the backaction
noise in Fig.\ \ref{Fig1} are uncorrelated, the ensemble dephasing rate 
is 
        \begin{equation}
\Gamma_\Sigma = (\Delta I)^2/4S_0 + A^2S_1/4\hbar^2 + \gamma_3
\label{GSigma}\end{equation}
(the same result can be obtained by translating Eq.\ (\ref{Bayes2-2})
into It\^o form and performing ensemble averaging).
Following Refs.\ \cite{Kor-LT,Kor-rev1}, let us characterize the
noise correlation by the magnitude (real number) 
\begin{equation}
K=\frac{S_{\varepsilon I}}{\hbar S_\Sigma},
\end{equation}
where $S_{\varepsilon I}$ is the mutual spectral density of the detector 
output noise and induced fluctuations of $\varepsilon$ (strictly speaking,
$S_{\varepsilon I}$ in our notation is only the real part of the 
mutual spectral density, while the imaginary part can formally describe 
the detector response \cite{Averin-book}). In our case 
$K=AS_1/\hbar S_\Sigma$. 
Expressing qubit evolution only in terms of physically relevant 
quantities, we obtain
        \begin{eqnarray} 
&&  \dot{\rho}_{11}=  -\dot{\rho}_{22}=  
-2\,\frac{H}{\hbar}\,\mbox{Im}\,\rho_{12}
         +\rho_{11}\rho_{22}\, \frac{2\Delta I}{S_\Sigma}\, [I(t)-I_0], 
        \label{Bayes1-3}\\ 
&& {\dot\rho}_{12}=  \imat\, \frac{\varepsilon}{\hbar }\,
        \rho_{12}+ 
        \imat \, \frac{H}{\hbar } \, (\rho_{11}-\rho_{22})
    + \imat K [I(t) -I_0] \, \rho_{12} 
\nonumber \\
&& \hspace{1.0cm}  -( \rho_{11}-  \rho_{22})  \frac{\Delta I}{S_\Sigma} \, 
[I(t)-I_0]\, \rho_{12} - \tilde\gamma  \rho_{12} ,
        \label{Bayes2-3}\end{eqnarray}
where 
\begin{equation}
\tilde\gamma = \Gamma_\Sigma - \frac{(\Delta I)^2}{4S_\Sigma} 
        - \frac{K^2S_\Sigma}{4}. 
\label{tgamma}\end{equation}

Eqs.\ (\ref{Bayes1-3})--(\ref{tgamma}) is the main result of this
Section. Comparing them with similar equations presented (but not derived)
in Refs.\ \cite{Kor-LT,Kor-rev1}, we notice a difference:
the term $\imat K[I(t)-I_0]\rho_{12}$ was incorrectly replaced in Refs.\ 
\cite{Kor-LT,Kor-rev1} by the term $\imat K[I(t)-(\rho_{11}I_1+\rho_{22}I_2)]
\rho_{12}$. 
Notice though that the effect of their difference 
$\imat K (\Delta I/2)(\rho_{11}-\rho_{22}) \rho_{12}$ is minor
since $\rho_{11}-\rho_{22}$ is the oscillating magnitude and averages to 
zero. As will be mentioned later, Eqs.\ (\ref{Bayes1-3})--(\ref{Bayes2-3})
in the case $\tilde\gamma  =0$ (this is possible for asymmetric ideal
detector) coincide with the corresponding equations of Ref.\ \cite{Goan2}
if $\varepsilon$ includes the detector-induced shift. 

        To translate Eqs.\ (\ref{Bayes1-3})--(\ref{Bayes2-3}) from
Stratonovich to It\^o form, notice that 
        \begin{equation} 
I(t)-I_0 = \frac{\Delta I}{2}\, (\rho_{11} -\rho_{22}) +\xi_0 (t) +\xi_1(t)
        \label{I(t)-2}\end{equation}
and the sum of two output white noises $\xi_{0+1}(t)\equiv\xi_0(t)+\xi_1(t)$ 
is the white noise with the spectral density $S_\Sigma$. Then using
the standard rule of translation, we obtain It\^o equations 
        \begin{eqnarray} 
&&  \dot{\rho}_{11}=  -\dot{\rho}_{22}=  
-2\,\frac{H}{\hbar}\,\mbox{Im}\,\rho_{12}
         +\rho_{11}\rho_{22}\, \frac{2\Delta I}{S_\Sigma}\, \xi_{0+1}(t), 
        \label{Bayes1-4}\\ 
&& {\dot\rho}_{12}=  \imat\, \frac{\varepsilon}{\hbar }\, 
        \rho_{12}+ 
        \imat \, \frac{H}{\hbar } \, (\rho_{11}-\rho_{22})
    + \imat K \xi_{0+1}(t) \, \rho_{12} 
\nonumber \\
&& \hspace{1.0cm}  -( \rho_{11}-  \rho_{22})  \frac{\Delta I}{S_\Sigma} \, 
\xi_{0+1}(t)\, \rho_{12} 
\nonumber \\
&& \hspace{1.0cm}  -  \left( 
\tilde\gamma + \frac{(\Delta I)^2}{4S_\Sigma}
+\frac{K^2S_\Sigma}{4}  \right) \rho_{12} , 
        \label{Bayes2-4}\end{eqnarray}
while the relation between output current $I(t)$ and pure noise
$\xi_{0+1}(t)$ is still given by Eq.\ (\ref{I(t)-2}). 
(Notice that the above mentioned incorrect term is correct in
the It\^o form of the equation, so the mistake was due to mixing of
the Stratonovich and It\^o forms.) The corresponding ensemble-averaged 
equations can be
obtained by erasing terms containing $\xi_{0+1}(t)$ in Eqs.\ 
(\ref{Bayes1-4})--(\ref{Bayes2-4}): 
        \begin{eqnarray}
&& 
   \dot{\rho}_{11}  =   -\dot{\rho}_{22}=  -2\,\frac{H}{\hbar}\,\mbox{Im}\,
          \rho_{12} \, ,  
        \label{aver1-2}\\ 
&&
 {\dot\rho}_{12}  =   \imat\, \frac{\varepsilon}{\hbar }\,\rho_{12} +   
        \imat \, \frac{H}{\hbar } \, (\rho_{11}-\rho_{22})
 - \Gamma_\Sigma  \, \rho_{12} \, .
        \label{aver2-2} 
        \end{eqnarray}
where the total ensemble decoherence rate $\Gamma_\Sigma$ is given by 
Eqs.\ (\ref{GSigma}) and/or (\ref{tgamma}).

        Since $\tilde\gamma >0$ [otherwise solution of Eqs.\ 
(\ref{Bayes1-4})--(\ref{Bayes2-4}) would violate inequality 
$|\rho_{12}|^2\leq \rho_{11}\rho_{22}]$, we obtain the fundamental 
limitation for the ensemble dephasing:
        \begin{equation}
\Gamma_{\Sigma} \geq \frac{(\Delta I)^2}{4S_\Sigma} +\frac{K^2S_\Sigma}{4}.
        \label{Gammalimit}\end{equation}
Besides the definition of the detector ideality $\eta$ given by Eq.\
(\ref{eta-gen}) which would give
$\eta = 1-(\tilde\gamma +K^2S_\Sigma /4)/\Gamma_\Sigma$, it is natural
to introduce another definition of the ideality\cite{Kor-rev1}
\begin{equation}
     \tilde\eta = 1-\frac{\tilde\gamma}{\Gamma_\Sigma}
        =\frac{(\Delta I)^2/4S_\Sigma +K^2S_\Sigma /4}{\Gamma_\Sigma},
\label{tildeeta}\end{equation}
since the term $K^2S_\Sigma /4$ does not correspond to the dephasing of
a single qubit. One more possible definition of ideality 
(which also gives 100\% if $\tilde\gamma =0$)  
is \cite{Kor-rev2}
        \begin{equation}
\tilde\eta_2 = \frac{1}{1+\tilde\gamma / [ (\Delta I)^2/4S_\Sigma ]}
=\frac{(\Delta I)^2/4S_\Sigma}{\Gamma_\Sigma - K^2S_\Sigma/4}, 
        \label{tildeeta2}\end{equation}
so that $(\tilde\eta_2)^{-1/2}$ directly corresponds\cite{Kor-rev2} 
to the total energy sensitivity of the detector in units of $\hbar /2$.
In the case $K=0$ all the definitions coincide, $\eta =\tilde\eta =
\tilde\eta_2$.

\section{Account of asymmetric ideal detector} 

        So far we have assumed that the ideal part of the detector 
in Fig.\ \ref{Fig1} does not induce the shift of the qubit energy 
asymmetry $\varepsilon$ (i.e.\ in our terminology assumed symmetrically 
coupled detector). However, in general the coupling with detector changes
$\varepsilon$, so it should be treated self-consistently.\cite{Kor-rev1}
As an example, the operating point of an SET as a detector is slightly
shifted by different charge states of a measured qubit. This generally 
leads to the change of the average potential $v$ of the SET island, which 
affects back the qubit energy asymmetry $\varepsilon$. Notice that 
$v$ can also be temporary shifted by a fluctuation of the current through
SET, leading to the correlation between the output and backaction 
noises. So, in this example the shift of $\varepsilon$ and noise correlation
are closely related. Similar situation occurs when as a detector we use
a QPC, which  location  relative  to  the  qubit  is  geometrically 
asymmetric.
\cite{Sprinzak,Kor-Av} Then the qubit state affects the phase of the QPC
current (of course, this should happen before the current becomes a classical
quantity), and in return each electron passing through the QPC affects the
phase difference between qubit states $|1\rangle$ and $|2\rangle$, thus
leading to effective shift of $\varepsilon$. Correspondingly, the noise 
of the QPC current causes the correlated noise of $\varepsilon$, so again 
these effects are closely related. 

        The asymmetric coupling can be relatively easy taken into account
for a small-transparency QPC using the model analyzed in Ref.\ \cite{Goan2}.
The detector and its interaction with the qubit are described by 
Hamiltonians 
       \begin{eqnarray}
&& {\cal H}_{det} = \sum_l E_l a_l^\dagger a_l +\sum_r E_r a_r^\dagger a_r 
+ \sum_{l,r} (T a_r^\dagger a_l+ T^\ast a_l^\dagger a_r),  
         \nonumber\\ 
&& {\cal H}_{int}= \sum_{l,r} 
(c_1^\dagger c_1 - c_2^\dagger c_2) 
\left( \frac{\Delta T}{2} \, a_r^\dagger a_l +
\frac{\Delta T^\ast}{2} \, a_l^\dagger a_r \right) , 
        \label{Hint}\end{eqnarray}
in which we neglect the dependence of tunnel matrix elements $T$ and 
$\Delta T$ on the electron states ($l,r$) in electrodes. The only 
difference of this model from the model of Ref.\ \cite{Gurvitz} 
[which leads to Eqs.\ (\ref{Bayes1})--(\ref{Bayes2})] is the possibility
of complex $T$ and $\Delta T$ (actually, $T$ can be assumed real
without loss of generality). 
Following the procedure developed in Ref.\ \cite{Gurvitz}, it is possible 
to show\cite{Rusko-unpub} that in the corresponding Bloch equation for 
$\dot\rho_{12}^n$ (where $\rho_{ij}^n$ is the density matrix 
with account of the number $n$ of electrons passed through the detector) 
the term \cite{Gurvitz} $\sqrt{I_1I_2}e^{-1}\rho_{12}^{n-1}$ should be 
replaced by $\exp(\imat \varphi ) \sqrt{I_1I_2}e^{-1}\rho_{12}^{n-1}$, 
where $\varphi = \arg[(T+\Delta T/2)(T^\ast -\Delta T^\ast /2)]$. 
Therefore, each electron tunneling through the detector adds the 
phase $\varphi$ 
to $\rho_{12}$, thus affecting the qubit energy asymmetry $\varepsilon$.

        The assumption of weak detector response implies 
$|\Delta T| \ll |T|$, so that $|\varphi | \ll 1$.\cite{Goan2} 
        The extra phase leads to the extra term 
$\imat ( \theta /\hbar ) I_d\rho_{12}$ in Eq.\ (\ref{Bayes2}) where 
$\theta \equiv \hbar \varphi /e$. Separating it into the average 
and fluctuating 
parts, we obtain the following equations for the asymmetric ideal detector 
in the Stratonovich form: 
        \begin{eqnarray}
&&  \dot{\rho}_{11}=  -\dot{\rho}_{22}=  
-2\,\frac{H}{\hbar}\,\mbox{Im}\,\rho_{12}
         +\rho_{11}\rho_{22}\, \frac{2\Delta I}{S_0}\, [I_d(t)-I_0], 
        \label{Bayes1-5}\\ 
&&  {\dot\rho}_{12}=  \imat\, \frac{\tilde\varepsilon}{\hbar }\,
        \rho_{12}+ 
        \imat \, \frac{H}{\hbar } \, (\rho_{11}-\rho_{22})
      + \imat \, \frac{\theta}{\hbar}\,  [I_d(t)-I_0] \, \rho_{12} 
\nonumber \\
&& \hspace{1cm}  -( \rho_{11}-  \rho_{22})  \frac{\Delta I}{S_0} \, 
[I_d(t)-I_0]\, \rho_{12} ,
        \label{Bayes2-5}\end{eqnarray}
where $\tilde\varepsilon = \varepsilon +\Delta \varepsilon$ and 
$\Delta\varepsilon = \theta I_0 = \hbar \varphi I_0/e$. Equations 
(\ref{Bayes1-5})--(\ref{Bayes2-5}) in It\^o form have been obtained 
in Ref.\ \cite{Goan2} (notice a different definition of $\theta$). 
\cite{Goan-notice} 
Let us mention that even though Eqs.\ 
(\ref{Bayes1-5})--(\ref{Bayes2-5}) have been derived for a particular 
model of the detector [Eq.\ (\ref{Hint})], it is expected 
that they are applicable to a significantly broader class of 
asymmetric ideal detectors (i.e.\ finite-transparency QPCs, quantum-limited
dc SQUIDs, etc.). Notice that Eqs.\ (\ref{Bayes1-5})--(\ref{Bayes2-5})
are formally similar to Eqs.\ (\ref{Bayes1-3})--(\ref{Bayes2-3}) 
except $\tilde\gamma =0$ (ideal detector) and $\varepsilon$ is replaced
by the self-consistent value $\tilde\varepsilon$. 

        The solution of Eqs.\ (\ref{Bayes1-5})--(\ref{Bayes2-5})
in the simple case $H=0$ is still given by Eqs.\ 
(\ref{QBayes1})--(\ref{QBayes2}) for $\rho_{11}$ and $\rho_{22}$,
while Eqs.\ (\ref{QBayes3})--(\ref{QBayes4}) for $\rho_{12}$ should be
replaced by
        \begin{eqnarray}
&& \rho_{12}(\tau ) = \rho_{12}(0) 
\frac{[\rho_{11}(\tau )\rho_{22}(\tau )]^{1/2}}
{[\rho_{11}(0)\rho_{22}(0)]^{1/2}}
\, e^{\imat \tilde\varepsilon \tau /\hbar} \, 
e^{\imat \theta (\overline{I_d}-I_0)\tau /\hbar}
        \label{QBayes3-2}\\
&& = \frac{
\rho_{12}(0) \exp [-\frac{(\Delta I)^2\tau}{4S_0} 
-\frac{(\overline{I_d}-I_0)^2\tau }{S_0}] 
\exp [\imat \, \frac{\tilde\varepsilon\tau +\theta (
\overline{I_d}-I_0) \tau }{\hbar}]
}{
\rho_{11}(0)\exp [-\frac{(\overline{I_d}-I_1)^2\tau }{S_0}]
+\rho_{22}(0)\exp [-\frac{(\overline{I_d}-I_2)^2\tau }{S_0}]
} .
        \label{QBayes4-2}\end{eqnarray} 

    Let us now add the classical noises $\xi_1$, $\xi_2$ and $\xi_3$
(see Fig.\ \ref{Fig1}) to our model of asymmetric ideal detector. 
Using the procedure explained in two previous Sections, 
we multiply Eq.\ (\ref{QBayes4-2}) by the factor 
$\exp (\imat Ax\tau /\hbar )$
where $x=\tau^{-1}\int_0^\tau \xi_1(t)dt$, average the resulting equation
and Eq.\ (\ref{QBayes1}) over the distribution $p(x)$ given by 
Eqs.\ (\ref{p1(x)})--(\ref{p2(x)}), differentiate the result
over time, and add the terms due to $H$ and $\xi_3$. In this way
we obtain the following Stratonovich equations:
        \begin{eqnarray} 
&&  \dot{\rho}_{11}=  -\dot{\rho}_{22}=  
-2\,\frac{H}{\hbar}\,\mbox{Im}\,\rho_{12}
         +\rho_{11}\rho_{22}\, \frac{2\Delta I}{S_\Sigma}\, [I(t)-I_0], 
        \label{Bayes1-6}\\ 
&& {\dot\rho}_{12}=  \imat\, \frac{\tilde\varepsilon}{\hbar }\,
        \rho_{12}+ 
        \imat \, \frac{H}{\hbar } \, (\rho_{11}-\rho_{22})
\nonumber\\
 &&\hspace{1.0cm} 
    + \frac{\imat}{\hbar}\,\frac{AS_1+\theta S_0}{S_\Sigma} \, 
 [I(t) -I_0] \, \rho_{12}  
\nonumber \\ 
&& \hspace{1.0cm}  
-( \rho_{11}-  \rho_{22})  \frac{\Delta I}{S_\Sigma} \, 
[I(t)-I_0]\, \rho_{12} 
 - \tilde\gamma  \rho_{12} ,
        \label{Bayes2-6}\end{eqnarray}
where 
\begin{equation}
\tilde\gamma = \frac{(\Delta I)^2S_1}{4S_0S_\Sigma} + 
\frac{S_0S_1}{S_\Sigma}\frac{(A-\theta )^2}{4\hbar^2} +\gamma_3 .
\end{equation}
It is interesting to notice that the single-qubit decoherence rate
$\tilde\gamma$ can be decreased by adding the backaction noise $\xi_2(t)
=A\xi_1(t)$ 
with $A=\theta$ (i.e.\ having the same correlation with the output noise 
as the quantum backaction), even though it increases the ensemble 
dephasing rate 
        \begin{equation}
\Gamma_\Sigma = (\Delta I)^2/4S_0 +\theta^2S_0/4\hbar^2 + 
A^2S_1/4\hbar^2 + \gamma_3.  
        \end{equation} 

        Introducing the total correlation between the output noise
and the backaction (including quantum) noise,
        \begin{equation}
K= \frac{AS_1+\theta S_0}{\hbar S_\Sigma} \, , 
        \label{K-def-2}\end{equation}
we reduce Eqs.\ (\ref{Bayes1-6})--(\ref{Bayes2-6}) to Eqs.\ 
(\ref{Bayes1-3})--(\ref{Bayes2-3}) with the only substitution of
$\varepsilon$ by the self-consistent value $\tilde\varepsilon$. 
Therefore, the corresponding It\^o equations are still  given 
by Eqs.\ (\ref{Bayes1-4})--(\ref{Bayes2-4}) and the ensemble-averaged
equations are still given by Eqs.\ (\ref{aver1-2})--(\ref{aver2-2}) 
(with $\varepsilon \rightarrow \tilde\varepsilon$), while the relation
between the single-qubit decoherence $\tilde\gamma$ and ensemble
decoherence $\Gamma_\Sigma$ still satisfies Eq.\ (\ref{tgamma}). 
Thus, we conclude that the Bayesian description of the measurement
process given by Eqs.\ (\ref{Bayes1-3})--(\ref{aver2-2}), which is 
expressed in terms 
of the physically relevant quantities $K$, $S_\Sigma$, and $\Gamma_\Sigma$, 
remains valid in the case when the ideal part of the detector is
assumed asymmetrically coupled to the qubit. Similarly, the
limitation (\ref{Gammalimit}) for $\Gamma_\Sigma$ remains valid
and the definitions (\ref{tildeeta})--(\ref{tildeeta2}) of the
detector ideality can still be used. 

    Let us emphasize that for the Bayesian description of the 
continuous measurement of a single qubit, the detector in our model
is completely described by six quantities: dc output current (``operating 
point'') $I_0$, response $\Delta I$, output noise $S_\Sigma$, 
ensemble dephasing rate $\Gamma_\Sigma$, correlation magnitude $K$, 
and the induced qubit energy shift $\Delta \varepsilon$
(the shift $\Delta \varepsilon$ can also have a classical contribution
due to shift of the detector operating point). 
The quantities $S_\Sigma$, $\Gamma_\Sigma$, and $K$ 
are analogous to the output, input, and cross-correlation 
noise terms, which are usually used for the description of a classical 
amplifier (and similarly for the description of a quantum amplifier --
see, e.g.\ Ref.\ \cite{Averin-book} and references therein). 
The induced energy shift $\Delta \varepsilon$ is somewhat similar to the
effect of a finite amplifier input impedance  onto the previous stage 
parameters.

        \section{Detector measuring entangled qubits} 

        Finally, let us consider the case when the detector is coupled
to $N$ arbitrarily entangled and arbitrary interacting qubits. 
Following Ref.\ \cite{Kor-ent} 
we introduce the measurement basis consisting of $2^N$ states $|i\rangle$
and up to $2^N$ different levels $I_i$ of the detector average current
(some average currents can coincide). For an ideal symmetric detector 
and ``frozen'' qubits, $H_{qbs}=0$, where $H_{qbs}$ is the Hamiltonian 
of intrinsic evolution of the qubits, the evolution of the density matrix
$\rho$ of qubits due to measurement 
is described by simple ``Quantum Bayes'' equation \cite{Gardiner,Kor-ent}
        \begin{equation}
\rho_{ij}(\tau ) =\rho_{ij}(0) \, 
        \frac{\sqrt{P_i(\tau )P_j(\tau)}}
     {\sum_k \rho_{kk}(0)P_k(\tau )},
        \label{QBayes-ent}\end{equation}
where $P_i(\tau )$ is the probability of obtaining particular 
measurement result ($\overline{I_d}$ in this case) assuming state 
$|i\rangle$ of the measured system: 
\begin{equation}
P_i(\tau )=\frac{1}{\sqrt{2\pi D_0}} 
\exp[-\frac{(\overline{I_d}-I_i)^2}{2D_0}]\, , \,\,\,
D_0=\frac{S_0}{2\tau} \, 
\end{equation}
(we again assume that the currents $I_i$ do not differ much, and therefore
the detector noise $S_0$ is state-independent). 

   To take into account additional output noise $\xi_1(t)$ with spectral
density $S_1$, we have to perform the averaging of the density matrix
$\rho_{ij}$ over $x=\overline{I}-\overline{I_d}$ with the weight factor 
\begin{eqnarray}
&& p(x)= P(x)/\int P(x')dx',
\label{p1(x)-2}\\
&& P(x)=\frac{\exp (-\frac{x^2}{2D_1}) }{\sqrt{2\pi D_1}} 
\sum_k \rho_{kk}(0)\frac{\exp [-\frac{(\overline{I}-x-I_k)^2}{2D_0}]}
{\sqrt{2\pi D_0}}
\label{p2(x)-2}\end{eqnarray}
This procedure (without account of other noise sources) will lead to 
the equation presented in Ref.\ \cite{Kor-ent} and corresponds to the
detector ideality $\eta =S_0/(S_0+S_1)$, similar to the one-qubit case. 

        For the classical backaction noise which is 100\% correlated 
with $\xi_1(t)$, we should take into account that 
it can be coupled differently to different qubits. Let us assume 
that the energy of each state $|i\rangle$ is affected by the classical 
backaction noise, proportional to $\xi_1$, so that $\varepsilon_i\rightarrow
\varepsilon_i + A_i\xi_1(t)$, where $A_i$ are arbitrary constants. 
Then Eq.\ (\ref{QBayes-ent}) should be multiplied by the factor 
$\exp [\imat (A_j-A_i)x\tau /\hbar]$ 
(in the one-qubit case the previously defined $A$ would correspond to
$A_2-A_1$). 

   Similarly, to take into account the possible asymmetry of the quantum
backaction noise, let us assume that each electron tunneling through
the ideal part of the detector shifts the phases corresponding to
states $|i\rangle$ (differently for different states), that leads 
to the extra factor 
$\exp [\imat (\theta_i -\theta_j) \overline{I_d}\tau /\hbar ]$ 
in Eq.\ (\ref{QBayes-ent}). 

    The uncorrelated classical noise $\xi_3(t)$ is also assumed to
be coupled differently to the states $|i\rangle$, so that 
$\varepsilon_i \rightarrow \varepsilon_i +g_i\xi_3(t)$, where 
$g_i$ are some constants. The averaging over noise $\xi_3$ is simple
and leads to the extra factor $\exp [-(g_i-g_j)^2S_3\tau /4\hbar^2]$
in Eq.\ (\ref{QBayes-ent}). 

        Taking into account the effect of $\theta_i$, averaging over 
the noise $\xi_1(t)$ (and fully correlated backaction noise) and $\xi_3(t)$,
differentiating equations over time, and adding the intrinsic evolution 
of qubits, we finally obtain the following equation in the Stratonovich
form:
        \begin{eqnarray}
&& \dot\rho_{ij} = -\frac{\imat}{\hbar} [H_{qbs},\rho]_{ij}
+ \imat \frac{\Delta \varepsilon_{ij}}{\hbar} \rho_{ij}
+ \imat K_{ij} [I(t)-\frac{I_i+I_j}{2}]\, \rho_{ij} 
\nonumber \\
 && \hspace{0.8cm}
 + \rho_{ij} \frac{1}{S_\Sigma} \sum_k \rho_{kk}
\left[ [I(t)-\frac{I_i+I_k}{2}](I_i-I_k)  \right. 
\nonumber\\
  && \hspace{1.2cm}
   \left. + [I(t)-\frac{I_j+I_k}{2}](I_j-I_k) \right] - 
        \tilde\gamma_{ij}\rho_{ij}, 
        \label{ent-Strat}\end{eqnarray}
where the first term describes the intrinsic evolution of qubits due
to $H_{qbs}$, $\Delta\varepsilon_{ij}=(\theta_i-\theta_j)(I_i+I_j)/2$ 
is the effective energy shift due to detector asymmetry, 
$S_\Sigma =S_0+S_1$ is the total output noise, 
\begin{equation}
  K_{ij}=\frac{(A_j-A_i)S_1+(\theta_i-\theta_j)S_0}{\hbar S_\Sigma}
\end{equation}
is the correlation factor between output and backaction noises, 
and the dephasing rate is 
\begin{eqnarray}
&& \tilde\gamma_{ij} = \frac{(I_i-I_j)^2S_1}{4S_0S_\Sigma} 
+\frac{(g_i-g_j)^2S_3}{4\hbar^2}
\nonumber \\
&& \hspace{1.0cm} 
+\frac{S_0S_1}{4\hbar^2S_\Sigma} [(A_j-A_i)-(\theta_i-\theta_j)]^2 . 
\end{eqnarray}
Notice that there are obviously no dephasing terms for diagonal matrix 
elements. Also notice that Eq.\ (\ref{ent-Strat}) is applicable to
both linear and nonlinear detectors, including purely quadratic detectors,
\cite{Averin-nonlin} as long as the condition of weak response is satisfied.

        Translating Eq.\ (\ref{ent-Strat}) into It\^o form, we obtain 
        \begin{eqnarray}
&& \dot\rho_{ij} = -\frac{\imat}{\hbar} [H_{qbs},\rho]_{ij}
+ \imat \frac{\Delta \varepsilon_{ij}}{\hbar} \rho_{ij}
+ \imat K_{ij} [I(t)-\sum_k\rho_{kk}I_k]\, \rho_{ij} 
\nonumber \\
 && \hspace{0.8cm}
 + \rho_{ij} \frac{1}{S_\Sigma} [I(t)-\sum_k\rho_{kk}I_k] 
    (I_i+I_j-2\sum_k\rho_{kk}I_k)
\nonumber \\
 && \hspace{0.8cm}
        -\Gamma_{ij}\rho_{ij}, 
        \label{ent-Ito}\end{eqnarray}
where the ensemble dephasing $\Gamma_{ij}$ is related to the single system
dephasing $\tilde\gamma_{ij}$ as 
\begin{equation}
    \Gamma_{ij}= \frac{(I_i-I_j)^2}{4S_\Sigma}+\frac{K_{ij}^2S_\Sigma}{4}
        +  \tilde\gamma_{ij} 
\label{ent-gamma-Gamma}\end{equation}
and in our particular case is equal to
\begin{eqnarray}
&&     \Gamma_{ij}= \frac{(I_i-I_j)^2}{4S_0}
+\frac{(A_j-A_i)^2S_1}{4\hbar^2}
+\frac{(\theta_i-\theta_j)^2S_0}{4\hbar^2}
\nonumber \\
&&\hspace{1.0cm}
+\frac{(g_i-g_j)^2S_3}{4\hbar^2}. 
\end{eqnarray}

Since the combination $I(t)-\sum_k\rho_{kk}I_k$ in Eq.\ (\ref{ent-Ito})
is a pure noise because of the relation 
        \begin{equation}
I(t) =\sum_k\rho_{kk}I_k + \xi_0(t)+\xi_1(t), 
        \end{equation}
the ensemble averaged evolution is described by the reduced equation 
\begin{equation}
 \dot\rho_{ij} = -\frac{\imat}{\hbar} [H_{qbs},\rho]_{ij}
+ \imat \frac{\Delta \varepsilon_{ij}}{\hbar} \rho_{ij}
        -\Gamma_{ij}\rho_{ij} . 
\label{ent-aver}\end{equation}

      Because of the reciprocity, it is natural to assume that
the backaction couplings $A_j-A_i$, $\theta_i-\theta_i$, and
$g_i-g_j$ are proportional to the signal coupling $I_i-I_j$, so that
$A_j-A_i=(I_i-I_j)a$, $\,\theta_i-\theta_j =  (I_i-I_j)\Theta $,
and $g_i-g_j= (I_i-I_j)g$.
(Actually, this assumption implies detector linearity and also that 
all interactions with qubits occur 
via one ``port of entry''. It is not valid, for example, when 
several geometrical parts of the detector interact with qubits in different
ways.) 
 With this assumption the parameters 
$\Delta \varepsilon_{ij}$,  $K_{ij}$, 
$\tilde\gamma_{ij}$, and $\Gamma_{ij}$ used in 
 evolution equations (\ref{ent-Strat})
and (\ref{ent-Ito}) become
\begin{eqnarray}
&& \Delta \varepsilon_{ij} =\Theta (I_i^2-I_j^2)/2,
\\
&& K_{ij}=\frac{aS_1+\Theta S_0}{\hbar S_\Sigma}\, (I_i-I_j),
\\
&& \tilde\gamma_{ij}= (I_i-I_j)^2 [ \frac{S_1}{4S_0S_\Sigma}  
+\frac{(a-\Theta)^2S_0S_1}{4\hbar^2S_\Sigma} + \gamma_{3,n} ] ,
\\
&& \Gamma_{ij}= (I_i-I_j)^2 (\frac{1}{4S_0}+\frac{a^2S_1}{4\hbar^2}
   +\frac{\Theta^2 S_0}{4\hbar^2}+\gamma_{3,n}),
\end{eqnarray}
where $\gamma_{3,n}=gS_3/4\hbar^2$.
Notice that there will be no 
dephasing between states $|i\rangle$ and $|j\rangle$ if the detector
is equally coupled to these states, $I_i=I_j$.

     The detector ideality in this case can be characterized by a single 
number (or few numbers for different definitions), which does not depend 
on the state of the measured system. Extending the definitions
(\ref{eta-gen}), (\ref{tildeeta}), and (\ref{tildeeta2}) discussed 
in previous Sections, 
the detector ideality can be characterized by the parameter combinations 
 \begin{eqnarray}
&& \eta =\frac{1/4S_\Sigma}{\Gamma_{\Sigma , n}}\, ,  \,\,\,\,\, 
\tilde\eta 
=\frac{1/4S_\Sigma + K_n^2S_\Sigma /4} 
{\Gamma_{\Sigma , n}} \, , \,\,\,\, 
\nonumber \\
&& \tilde\eta_2 = \frac{1/4S_\Sigma}{\Gamma_{\Sigma ,n}-K_n^2S_\Sigma /4} 
\, , 
  \end{eqnarray}
where $K_n\equiv (aS_1+\Theta S_0)/\hbar S_\Sigma$ and 
$\Gamma_{\Sigma ,n} \equiv 
1/4S_0+a^2S_1/4\hbar^2+\Theta^2S_0/4\hbar^2+\gamma_{3,n}$.
In the case $a=\Theta =0$ all definitions 
of ideality coincide and 
the evolution equation (\ref{ent-Strat}) reduces to the equation
derived in Ref.\ \cite{Kor-ent}. In the case of 
finite $a$ and/or $\Theta$, more natural definitions are $\tilde\eta$ and 
$\tilde\eta_2$ (again, $\tilde\eta_2^{-1/2}$ is the total energy sensitivity
in units of $\hbar /2$). However, ideality $\eta$ can also be a useful
parameter, for example, if there is no way to control the degree of 
freedom affected by the backaction noise $\Theta \xi_0+a\xi_1$, 
and therefore the corresponding dephasing cannot be reduced by a feedback
procedure.

\section{Conclusion}

In this paper we have analyzed the process of continuous measurement 
of a solid state qubit by a nonideal solid state detector. 
We have considered the phenomenological model of the 
detector (Fig.\ \ref{Fig1}) consisting of an ideal (quantum-limited) 
part and classical noise sources which contribute to the output ($\xi_1$) 
and backaction ($\xi_2+\xi_3$) noises. The possible correlation between
classical output and backaction noise sources is taken into account by
separating the backaction noise into a contribution $\xi_2(t)$ fully 
correlated with output noise $\xi_1(t)$ and the uncorrelated contribution
$\xi_3(t)$. For the description of the ideal part we have started with
the Bayesian equations of Refs.\ \cite{Kor-99,Kor-rev2,Kor-rev1} and then 
used the model of an asymmetrically coupled ideal detector developed
in Ref.\ \cite{Goan2}. The asymmetric coupling changes the self-consistent 
energy difference between two qubit states. Also, this change fluctuates
in time and the fluctuations are correlated with the output noise, thus 
producing an effect similar to the correlation of classical noises. 

        The main result of the paper for the one-qubit case is the 
derivation of evolution equations (\ref{Bayes1-3})--(\ref{Bayes2-3}) and 
(\ref{Bayes1-4})--(\ref{Bayes2-4}) in Stratonovich and It\^o form,
respectively (the qubit energy parameter $\varepsilon$ should be
treated self-consistently, as discussed in section V). 
In these equations the detector is characterized by the  
total output noise $S_\Sigma$, induced ensemble qubit decoherence rate 
$\Gamma_\Sigma$, and the total correlation $K$ [see Eq.\ (\ref{K-def-2})] 
between output and backaction noises, so that the detector separation 
into the quantum part and extra noises is irrelevant.
(Notice that these three quantities are the counterparts of output, input, 
and cross-correlation noise terms used for the description of a classical
amplifier; the induced qubit energy shift $\Delta \varepsilon$ is somewhat 
analogous to a backaction due to finite input impedance.) 
 The relation
between ensemble and single qubit decoherence rates
is given by Eq.\ (\ref{tgamma}), which leads to the fundamental
limitation (\ref{Gammalimit}) for the ensemble decoherence rate.
The discussed definitions of the detector ideality [see Eqs.\ 
(\ref{eta-gen}), (\ref{tildeeta}), and (\ref{tildeeta2})] are various
combinations of the single qubit decoherence rate, ensemble decoherence, 
and the ``information acquisition'' rate $(\Delta I)^2/4S_\Sigma$. A 100\% 
ideal detector corresponds to the absence of single qubit decoherence. 

    The theory developed for a single qubit measurement is generalized
to the case of entangled qubits in Section VI. The evolution equation
is given by Eqs.\ (\ref{ent-Strat}) and (\ref{ent-Ito}), while the
relation between ensemble and single system decoherence rates is given
by Eq.\ (\ref{ent-gamma-Gamma}).

\vspace{0.5cm}

The author thanks R. Ruskov for permission to use his result on ideal
asymmetric detector prior to publication and 
 for critical reading of the manuscript. 
The work was supported by NSA and ARDA under ARO grant DAAD19-01-1-0491.



\begin{references}
\bibitem{symmetric} Assumption of a symmetric detector means that 
        the measurement process does not shift the energy levels of
        the measured qubit.
\bibitem{Goan-notice} In Ref.\ \cite{Goan2} it is stated that the effect
        of finite $\theta$ is always small in the ``quantum diffusion''
        (``weakly responding'') case, $|\Delta I|\ll I_0$. We think that 
        actually its effect can be significant.
\end{references}
\end{document}